\documentclass[aps,twocolumn,showpacs,preprintnumbers,nofootinbib,superscriptaddress]{revtex4-1}

\usepackage{amsmath,amssymb,bm}
\usepackage[dvips]{graphicx}
\usepackage{hyperref}
\usepackage{yfonts}
\usepackage{color}
\newcommand{\beq}{\begin{eqnarray}}
\newcommand{\eeq}{\end{eqnarray}}
\newcommand{\tr}{\text{tr}}
\newcommand{\SU}{\text{SU}}
\newcommand{\U}{\text{U}}
\newcommand{\UA}{\U(1)_{\rm A}}
\renewcommand{\d}{\partial}
\newcommand{\LL}{\mathcal{A}}
\begin{document}

\title{Quasi-instantons in QCD with chiral symmetry restoration}

\author{Takuya Kanazawa}
\affiliation{iTHES Research Group and Quantum Hadron Physics Laboratory, 
RIKEN, Wako, Saitama 351-0198, Japan}

\author{Naoki Yamamoto}
\affiliation{Department of Physics, Keio University,
Yokohama 223-8522, Japan}

\preprint{RIKEN-QHP-168}

\begin{abstract}
We show, without using semiclassical approximations, that, in 
high-temperature QCD with chiral symmetry restoration and $\UA$ symmetry breaking, 
the partition function for sufficiently light quarks can be expressed as an ensemble 
of noninteracting objects with topological charge that obey the Poisson statistics. 
We argue that the topological objects are ``quasi-instantons'' (rather than bare instantons) 
taking into account quantum effects. Our result is valid even close 
to the (pseudo)critical temperature of the chiral phase transition.
\end{abstract}
\maketitle

\section{Introduction}
The instantons \cite{Gross:1980br, Schafer:1996wv}, which are topological excitations of 
QCD in Euclidean spacetime, play key roles in nonperturbative aspects of QCD. In particular, 
they solve the $\UA$ puzzle \cite{'tHooft:1976up,'tHooft:1986nc} and provide us with a 
semiclassical picture of chiral symmetry breaking in the QCD vacuum \cite{Schafer:1996wv}.
The instanton liquid model \cite{Schafer:1996wv} is proposed to describe 
the physically relevant regimes at finite temperature and density that are accessible 
in heavy ion collision experiments at RHIC/LHC. However, model-independent 
many-body instanton descriptions based on QCD are reliable only in some extreme 
cases: QCD at sufficiently high temperature, $T \gg T_c$ with $T_c$ being the 
(pseudo)critical temperature of chiral transition \cite{Gross:1980br,Khoze:1990nt,Dunne:2010gd}, 
or at sufficiently large chemical potential, $\mu \gg \Lambda_{\rm QCD}$ 
\cite{Son:2001jm, Schafer:2002ty, Yamamoto:2008zw}. 
This is thanks to the presence of small expansion parameters,  
$\Lambda_{\rm QCD}/T \ll 1$ and $\Lambda_{\rm QCD}/\mu \ll 1$, but it is 
not the case for $T \sim \Lambda_{\rm QCD}$ and/or $\mu \sim \Lambda_{\rm QCD}$.
This is one of the obstacles that prevents us from understanding the properties 
of physically interesting phases at finite $T$ and/or $\mu$ on the basis of QCD itself.

In this paper, we show that QCD at $T>T_c$, even close to $T_c$,
can be seen as an ensemble of \emph{noninteracting quasi-instantons}, 
which could be seen as ``dressed'' quasiparticles in the strongly interacting 
ensemble of conventional instantons (calorons). 
We demonstrate this picture in a theoretically controlled manner as follows%
\footnote{We note that the main assumption in this paper 
is the breaking of $\UA$ symmetry through the axial anomaly 
[or precisely speaking, $f_A \neq 0$ in Eq.~(\ref{f})].
This is shown in QCD at $T \gg T_c$ in Refs.~\cite{Gross:1980br,Khoze:1990nt,Dunne:2010gd} 
through controlled semiclassical calculations (see also Ref.~\cite{Laine:2003bd}),  
and it seems natural to expect $f_A \neq 0$ at \emph{any} $T>T_c$ 
(see, however, Refs.~\cite{Shuryak:1993ee, Cohen:1996ng, Lee:1996zy, Evans:1996wf, Aoki:2012yj}).}:
we start with the effective theory in terms of the quark mass $M$ that 
includes the dependence on the $\theta$ angle and quantum fluctuations 
systematically. We then show that the partition function of this effective theory 
can be understood as an ensemble of noninteracting objects with positive or 
negative topological charges which obey the Poisson distribution. We shall 
refer to such objects as \emph{(anti-)quasi-instantons}.

Our approach should be contrasted with the conventional instanton analysis 
where one adds quantum effects to the classical instantons step by step. 
In our approach, quantum effects are already included in the effective 
theory, and then we rewrite it in a physically different form in terms of the 
quasi-instantons. In particular, we do \emph{not} use the semiclassical 
approximation in this approach. 
While this procedure looks similar to the ones \cite{Son:2001jm,Yamamoto:2008zw} 
considered in the color-superconducting phases of QCD at zero temperature 
and high density, the starting effective theories and expansion
parameters are completely different.

For simplicity, we first consider $N_f=2$ as an example, and later extend it to 
general $N_f \geq 2$. Our argument is also applicable to QCD at finite chemical 
potential $\mu$ and/or in a magnetic field $B$ in a straightforward manner, 
as long as the chiral symmetry of QCD, except $\UA$, is restored.

\section{QCD partition function}
Let us consider the QCD partition function as a function of quark mass $M$. 
Here we consider the high-temperature regime $T>T_c$ and expand the 
free energy of QCD in terms of a small parameter 
$m/T \lesssim m/\Lambda_{\rm QCD} \ll 1$.
Recalling that quarks acquire the effective thermal mass and
there are no massless Nambu-Goldstone modes at $T > T_c$,
the free energy of the system is analytic in quark mass.

In order to write down the general form of the free energy in terms
of $M$, we use the spurion field method. We allow the quark mass 
matrix $M$ to transform under the symmetry,
${\cal G} \equiv \SU(N_f)_{\rm R} \times \SU(N_f)_{\rm L} \times \UA$, 
so that the mass term in the original QCD Lagrangian,
\beq
  {\cal L}_{\rm mass} = \psi_{L}^{\dag} M \psi_R + {\rm h.c.},
\eeq
is invariant under ${\cal G}$, where $\psi_{R,L}$ are the right- and left-handed quarks.
As $\psi_{R, L}$ transform under ${\cal G}$ as 
\beq
  \psi_R \rightarrow e^{i\theta_A} V_R \psi_R, \quad 
  \psi_L \rightarrow e^{-i\theta_A} V_L \psi_L,
\eeq
with $\theta_A$ being the $\UA$ phase, we impose the following
transformation law for $M$:
\beq
  \label{M}
  M \rightarrow e^{-2 i\theta_A} V_L M V_R^{\dag}.
\eeq

Using Eq.~(\ref{M}), one can construct the general free energy density of 
two-flavor QCD invariant under $\SU(2)_{\rm R} \times \SU(2)_{\rm L}$ 
but \emph{not} under $\UA$, as 
\begin{align}
  \label{f}
  f = f_0 -  f_2 \tr M^{\dag} M  - f_A (\det M + \det M^{\dag}) + O(M^4)\,.
\end{align}
Here $f_0$, $f_2$, and $f_A$ are parameters that are functions of 
$T$ and $V_3$ (the spatial volume) \cite{reg}.
The term $i(\det M-\det M^\dagger)$, 
which breaks parity ($M\leftrightarrow M^\dagger$), is omitted.
While the $f_2$ term is chirally symmetric and 
invariant under $\UA$, the $f_A$ term is \emph{not} invariant under $\UA$ 
and expresses the QCD axial anomaly. 
One microscopic explanation for the $f_A$ term may be given by the 
(bare) instanton-induced interactions \cite{'tHooft:1976up}, but 
our discussion below does not depend on the specific origin of $f_A$; 
the main assumption we make is $f_A \neq 0$.

Superficially, Eq.~(\ref{f}) might look similar to the Ginzburg-Landau theory, where 
the free energy is expanded in terms of a small order parameter around a phase 
transition based on symmetries. In Eq.~(\ref{f}), on the other hand, high-order terms 
in $M$ are suppressed in terms of the small parameter $m/T \ll 1$, and the 
expansion is \emph{not} limited to the region near the chiral transition; also,
$M$ is not an order parameter for some symmetry.%
\footnote{\label{fn2}%
In the case of a second-order chiral phase transition, 
one needs to let $m\rightarrow 0$ in the limit $T\rightarrow T_c$ 
in order to keep the mass expansion \eqref{f} convergent. This is because 
the coefficients in Eq.~(\ref{f}) tend to diverge as $T\rightarrow T_c$, which 
reflects the nonanalytic mass dependence of $f$ at $T=T_c$. 
}

\section{The $\theta$ angle and topological susceptibility}
One can incorporate the dependence of the $\theta$ angle in Eq.~(\ref{f}) 
via the replacement, $M \rightarrow M e^{i\theta/N_f}$.
This is because, for 
\beq
  Z(\theta) = \sum_{q=-\infty}^{\infty} e^{iq \theta} Z_q,
\eeq
the zero-mode contribution to the fermion determinant in $Z_q$ is 
given by $(\det M)^q$ for $q\geq 0$ and $(\det M^\dagger)^{-q}$ for $q<0$, 
and so $Z(\theta)$ depends on $\theta$ only
through the combination $M e^{i \theta/N_f}$ \cite{Leutwyler:1992yt}.
In two-flavor QCD with $M = {\rm diag} (m_u, m_d)$, the resulting free energy is 
\beq
  \label{f_tilde}
  f(\theta) = \tilde f - 2 f_A m_u m_d \cos \theta + O(m^4),
\eeq
where we define $\tilde f = f_0 - f_2 (m_u^2 + m_d^2)$.

With this free energy density $f(\theta)$, the partition function is given by
\beq
  \label{Z_QCD}
  Z_{\rm QCD}(\theta) = \exp[{- V_4 f(\theta)}],
\eeq
where $V_4 \equiv V_3/T$ is the four-volume.
It is then easy to derive the topological susceptibility, 
\beq
  \chi \equiv - \frac{1}{V_4}
  \left. \frac{\d^2 \ln Z_{\rm QCD}}{\d \theta^2} \right|_{\theta=0} = 2f_A m_u m_d 
  + O(m^4).
  \label{chi}
\eeq
%To the best of our knowledge, this is a new result.
This should be contrasted with the topological susceptibility in the QCD vacuum, 
$\chi_{\rm vac} = \Sigma (m_u^{-1} + m_d^{-1})^{-1}$, with $\Sigma$ being the 
magnitude of the chiral condensate \cite{Leutwyler:1992yt}. 
This difference is because the free energy of the QCD vacuum 
has an $O(M)$ term due to the spontaneous chiral symmetry breaking, 
whereas the free energy at $T>T_c$ does not [see Eq.~(\ref{f})]. 

Equation~\eqref{chi} is consistent with the anomalous Ward identity derived by 
Veneziano \cite{Veneziano:1979ec}, which for general $N_f$ and 
for equal masses states that
\begin{align}
  \chi & = - \frac{m}{N_f^2}\langle\bar\psi_f\psi_f\rangle 
  + \frac{m^2}{N_f^2}\int d^4x \,\langle 
  \bar\psi_f\gamma_5\psi_f(x)\bar\psi_g\gamma_5\psi_g(0) \rangle\,,
  \label{aWI}
\end{align} 
where the sum over repeated indices is understood ($f,g=1,\dots,N_f$). 
Indeed, for two-flavor QCD ($m_u=m_d=m$),
substituting $\langle\bar\psi_f\psi_f\rangle=-4(f_2+f_A)m+O(m^3)$ and 
$\int d^4x\langle \bar\psi_f\gamma_5\psi_f(x)\bar\psi_g\gamma_5\psi_g(0) \rangle
= 4(f_A-f_2)+O(m^2)$ that follow from Eq.~\eqref{f} into the rhs of Eq.~\eqref{aWI}, 
one gets $\chi=2 f_A m^2+O(m^4)$, in agreement with Eq.~\eqref{chi}. However, 
our direct derivation of the simple expression \eqref{chi} from the expansion of 
the QCD free energy at $T>T_c$ is new. Within our approach, not only the topological 
susceptibility but also the higher-order moments of the winding number $q$ can be 
derived in a straightforward manner, by taking derivatives of $\ln Z_{\rm QCD}$ with 
$\theta$ at $\theta=0$ as
\begin{subequations}\label{higher}
\begin{align}
  \langle q^2 \rangle  & = \LL,
  \\
  \langle q^4 \rangle  & = \LL(1+3\LL),
  \\
  \langle q^6 \rangle  & = \LL(1+15\LL+15\LL^2),
  \\
  \langle q^8 \rangle  & = \LL(1+63\LL+210\LL^2+105\LL^3),
\end{align}
\end{subequations}
with $\LL\equiv 2V_4 f_Am^{2}$, while all odd moments vanish.
(We will derive the results for general $N_f$ later.)
If $\LL$ is fixed in the limit $m\to 0$ and $V_4\to\infty$, then all 
the other $\theta$-dependent terms in $\ln Z_{\rm QCD}(\theta)$ 
drop off and these results become \emph{exact}.  

We emphasize that the arguments above are based only on the symmetry and 
analyticity of QCD under the systematic expansion, and thus are fully under theoretical 
control. Note that, while analytical calculations of the $\theta$-dependence are also possible 
at sufficiently high $T\gg\Lambda_{\rm QCD}$ based on a dilute instanton gas picture 
\cite{Gross:1980br}, in the present paper we do not rely on the assumption 
of high $T$; indeed, the above arguments are valid at generic $T>T_c$.  
Although the radius of convergence of the expansion \eqref{f} may vary 
with $T$, it does not affect our main results.

Equation~(\ref{Z_QCD}) can be expressed as 
\begin{align}
  Z_{\rm QCD} & = e^{-V_4 \tilde f} 
  \exp \bigg (V_4 \lambda \sum_{Q={\pm 1}}e^{iQ\theta} \bigg)
  \nonumber 
  \\
  & = e^{-V_4 \tilde f} \sum_{N=0}^{\infty} \frac{(V_4 \lambda)^N}{N!}
  \bigg( \sum_{Q={\pm 1}}e^{iQ\theta} \bigg)^{\! \! N},
\end{align}
where $\lambda \equiv f_A m_u m_d$ and we used the Taylor expansion
in the second line above. The $O(m^4)$ contribution is small and is 
disregarded in this expression. By using the identity
\begin{align}
  \bigg(\sum_{Q={\pm 1}}e^{iQ\theta} \bigg)^{\! \! N} 
  & = \sum_{Q_1=\pm 1}\cdots\sum_{Q_N=\pm 1}
  \left(e^{iQ_1\theta}\cdots e^{iQ_N\theta}\right)
  \notag
  \\
  & = 
  \sum_{N_+ + N_- = N}\frac{N!}{N_+! N_-!}\,  e^{i q \theta}\, ,
  \label{identity}
\end{align}
where $q \equiv \sum_{i=1}^N Q_i = N_+ - N_-$ with $N_\pm\geq 0$, 
we arrive at the expression
\begin{align}
  \label{Z_inst}
  Z_{\rm QCD} &= e^{-V_4 \tilde f}  
  \sum_{N_{+} = 0}^{\infty}
  \frac{(V_4 \lambda \,e^{i\theta})^{N_+}}{N_+ !}
  \sum_{N_{-} = 0}^{\infty}
  \frac{(V_4 \lambda \,e^{-i\theta})^{N_-}}{N_- !}
  \,. 
\end{align}
This is the main equation of this paper. 
We will denote the piece in the summation in Eq.~(\ref{Z_inst}) 
as $g(N_{\pm}, \theta)$ for later convenience.

\section{Quasi-instanton ensemble interpretation}
To understand the physical meaning of Eq.~(\ref{Z_inst}), first recall that
the $\theta$ angle enters the original QCD action only through 
the combination
\beq
  S_{\theta} = - i \theta Q_{\rm top}, \quad 
  Q_{\rm top} \equiv \frac{1}{32\pi^2} \int d^4x\, F^a_{\mu \nu} \tilde F^a_{\mu \nu}.
\eeq
In Eq.~(\ref{identity}), on the other hand, the $\theta$ angle appears
through the combination $i q\theta$, so one can make the identification  
$q = Q_{\rm top}$. One can now understand that each $Q_i = \pm 1$
is to be identified with the integer topological charge. Hence Eq.~(\ref{Z_inst})
means that the QCD partition function can be written as an ensemble of $N_{\pm}$
\emph{noninteracting} objects that have positive or negative topological charges 
$Q_i=\pm 1$. 
We shall call them the (anti-)quasi-instantons, as they are different from the bare 
(anti-)instantons in that they are dressed with the classical and quantum effects 
of interactions. 

This result may seem at first sight trivial due to the following reason: for a sufficiently 
small mass $m\ll T$, the statistical weight of an isolated instanton $\propto m^{N_f}$ is suppressed 
and instantons become dilute. Then the average distance between instantons is so large that  
they may well behave as effectively noninteracting objects. This picture is oversimplified, however, 
for the actual gauge field is not a dilute instanton gas (unless $T\gg T_c$), but rather 
a crowded mixture of (anti-)instantons overlapping with each other. In particular, 
the so-called instanton--anti-instanton molecules \cite{Schafer:1996wv}, which 
are topologically neutral and evade suppression by powers of $m$, will proliferate and 
call for a treatment as a strongly coupled many-body system. 
Therefore, it is a highly nontrivial finding that, despite nontrivial interactions between bare 
instantons (see, e.g., Ref.~\cite{Schafer:1996wv}),
after integrating out the effects of interactions [which is indirectly done through 
Eq.~(\ref{f}), where the effects of interactions are included in the coefficients of the 
expansion], the resulting quasi-instantons turn out to be noninteracting. 
This is somewhat similar to the idea of Landau's Fermi liquid theory \cite{Landau}, 
where the system is described in terms of weakly interacting quasiparticles that have the 
same quantum numbers as the original particles but are dressed with the effects of interactions.

A similar identification of ``quasi-instantons'' in the two-flavor color superconductivity 
and in the color-flavor locked phase of QCD at high density were previously noted in 
Refs.~\cite{Son:2001jm} and \cite{Yamamoto:2008zw}, respectively. 
There, they arrive at the ensemble of \emph{weakly interacting} quasi-instantons 
through a different path: the duality mapping of the low-energy effective theories 
for the $\eta$ and $\eta'$ mesons, respectively. In contrast, the quasi-instantons 
here do not interact with each other, unlike Refs.~\cite{Son:2001jm, Yamamoto:2008zw}. 
Physically, this difference is due to the fact that in hot QCD, there is no (pseudo-)Nambu-Goldstone 
mode associated with some symmetry breaking that mediates the interaction between 
(anti-)quasi-instantons in the present case.

\section{Moments and Poisson distribution}
Once the partition function is understood in terms of (anti-)quasi-instantons, it is easy 
to derive the density of (anti-)quasi-instantons, susceptibility, and higher moments,
following Ref.~\cite{Yamamoto:2008zw}. For example, at $\theta=0$ we have
\beq
  \label{eq:Npm}
  \langle N_{\pm} \rangle = \frac{\exp(-V_4 \tilde f) }{Z_{\rm QCD}} \sum_{N_{\pm} = 0}^{\infty} 
  N_{\pm} g(N_{\pm}, 0) = V_4 \lambda,
\eeq
with $\langle\dots\rangle$ the expectation value with respect to the QCD measure,
and so the number density of (anti-)quasi-instantons is 
$n_{\pm} = \langle N_+ \rangle/V_4 = \lambda =f_A m_u m_d$.
Therefore, the average topological charge is $\langle Q \rangle= 0$, and the 
total quasi-instanton density is $n \equiv n_+ + n_- = 2 \lambda$;
the average spatial distance between quasi-instantons is
$d_q\sim (f_A m^2/T)^{-1/3}$.

By a procedure similar to the above, one can show \cite{Yamamoto:2008zw}
\beq
\label{moment}
  \left \langle \frac{N_+ !}{(N_+ -k)!} \frac{N_- !}{(N_- -l)!}  \right \rangle
  = (\lambda V_4)^{k + l}
\eeq
for any non-negative integer $k$ and $l$.  
Recalling the property of Poisson statistics, 
$f(x) = e^{-\beta} {\beta^x}/{x!}$\,, that its $n$th factorial moment is $\beta^n$, 
we can conclude that the quasi-instantons and anti-quasi-instantons
follow the Poisson statistics independently. Usually, the Poisson 
distribution for the ensemble of \emph{bare} instantons and anti-instantons 
is expected only at $T \gg T_c$ \cite{Gross:1980br}.
Here, on the other hand, we have shown that the ensemble of \emph{quasi}-instantons 
always obey the Poisson distribution at $T > T_c$, even for $ T \sim T_c$,%
\footnote{%
The noninteracting quasi-instanton picture should remain valid  
if $d_q\gg \xi$, where $\xi$ is the correlation length of the system.
For a first-order transition, this condition is satisfied if $m\ll T$.
For a second-order transition, $\xi\sim(T-T_c)^{-\nu}$ diverges, 
so one has to judiciously let $m\rightarrow 0$ near $T_c$. 
Note that this requirement has already appeared in footnote \ref{fn2}. 
}
in a theoretically controlled manner. As $T \rightarrow \infty$,  
the quasi-instantons are expected to reduce to bare instantons, and  
$f_A$ is estimated to be proportional to the one-instanton weight, $e^{-8\pi^2/g(T)^2}\ll 1$, 
with $g(T)$ the gauge coupling at the scale $T$. 

Note that higher-order terms with $\theta$ dependence in Eq.~(\ref{f}), 
e.g., $(\det M)^2 e^{2i\theta}+\text{h.c.}$, are related to multi-instanton effects, 
but are suppressed by some powers of $m_f / T_c \ll 1$. 
If $m_f$ is not sufficiently small compared with $T_c$, 
such effects would become important and distort the Poisson distribution.

For the Poisson distribution, the susceptibility is equal to 
the first moment, and thus we have 
\beq
  \chi_{\rm inst} = n_{\rm inst} = 2 \lambda,
\eeq
which is equal to Eq.~(\ref{chi}). This is valid for any $T>T_c$.
From Eq.~(\ref{moment}), higher moments in Eq.~(\ref{higher}) can 
also be reproduced.

\section{Extension}
So far we have concentrated on $N_f = 2$. From now on, we generalize
our argument to any $N_f \geq 2$. In this case, we use Eq.~(\ref{M}) again, 
but we need to modify the form of the free energy in Eq.~(\ref{f}).
Considering all the possible terms to $O(M^{N_f})$ consistent with the 
symmetry ${\cal G}$, the free energy is given by
\beq
  f(\theta) = F(M)  - f_A (e^{i \theta} \det M + e^{-i\theta} \det M^{\dag}) + O(M^{N_f + 1}).
  \nonumber 
\eeq
Here the only term involving the $\theta$ angle is the $f_A$ term, and we 
denote all the other terms to $O(M^{N_f})$ by the $F(M)$ term.
For $M = {\rm diag}(m_1,m_2,\cdots,m_{N_f})$, the free energy reads 
\beq
  \label{F}
  f(\theta) = F(m_i) - 2 f_A \cos \theta \prod_{i=1}^{N_f} m_i \,,
\eeq
where $O(M^{N_f + 1})$ terms have been neglected. 
Following the same steps as before, one gets the expression 
\begin{align}
  \!\!\! 
  Z_{\rm QCD} &= \exp(-V_4 F)  \sum_{N_{\pm} = 0}^{\infty}
  \frac{(V_4 \sigma)^{N_+ + N_-}}{N_+ ! \, N_- !}e^{i \theta (N_+-N_-)},
\end{align}
where $\sigma \equiv f_A \prod_{i=1}^{N_f}m_i$. One can also repeat the 
same interpretation in terms of (anti-)quasi-instantons as above, 
with the replacements $\tilde f \rightarrow F$
and $\lambda \rightarrow \sigma$. In particular, we have for the 
quasi-instanton density and the topological susceptibility
\beq
  \chi_{\rm inst} = n_{\rm inst} = 2 \sigma.
\eeq
The higher moments are given in Eq.~(\ref{higher}) with the replacement 
$\LL \rightarrow 2 V_4 \sigma$.

Finally, we comment on one-flavor QCD. 
In one-flavor QCD vacuum, the $\theta$ dependence of the free energy is $\sim \cos \theta$
\cite{Leutwyler:1992yt}, similarly to Eqs.~(\ref{f_tilde}) and (\ref{F}). 
In this case, one can expand the free energy density in terms of $m/\Lambda_{\rm QCD} \ll 1$ as
\beq
  f_{N_f=1} = f_0 - \Sigma m \cos \theta + O(m^2),
\eeq
where $\Sigma$ is the magnitude of the chiral condensate.
One can reach the noninteracting quasi-instanton picture in this case as well.
This is not limited to high temperature but holds true for \emph{any} $T>0$ because, 
for $N_f = 1$, chiral symmetry is always broken explicitly by the $\UA$ anomaly \cite{Khoze:1990nt}.

\section{Conclusion}
From the systematic effective theory that includes the $\theta$ 
dependence and quantum effects of QCD at $T > T_c$, we arrived 
at the simple picture of noninteracting (anti-)quasi-instantons for
sufficiently light quarks. 
This picture may explain why the dilute instanton gas approximation, which 
has been conventionally expected to be valid only at sufficiently high temperature 
$T \gg T_c$\,, provides a reasonable description of QCD even close to $T_c$\,, 
as observed in lattice QCD simulations \cite{Edwards:1999zm, Buchoff:2013nra} 
(see, however, Ref.~\cite{Aoki:2012yj}). A similar instanton gas behavior was also 
numerically observed for the pure $\SU(N_c)$ (QCD with $m = \infty$) and $G_2$ 
lattice gauge theories just above $T_c$ \cite{Bonati:2013tt,Bonati:2015uga}. 
This might imply that the noninteracting instanton picture at $T>T_c$ is valid 
independently of $m$, though it can be theoretically justified only for small $m$ 
from our argument at this moment. 

Note that our argument above does not explicitly depend on $T$, 
except that chiral symmetry is restored at $T>T_c$.
This implies that our argument and results are also applicable to dense 
and magnetized matter. The coefficients of the expansion in $M$ then 
become functions of $T$, $\mu$, and $B$ \cite{B}. 
We also stress that our argument so far works only in a chirally symmetric 
phase of QCD. In a phase with spontaneous chiral symmetry breaking,
the $\theta$ dependence of the free energy is not like Eqs.~(\ref{f_tilde})
and (\ref{F}) \cite{Leutwyler:1992yt}. 

It would be an interesting question to study various correlation functions
at $T > T_c$ in terms of the quasi-instanton picture. One may also be able to 
understand the spectral density of the Dirac operator in relation to (the breaking of) 
the $\UA$ symmetry at $T>T_c$ in terms of the quasi-instanton picture.
We will defer answering such questions to future work  \cite{Kanazawa}.

\vspace{0.7cm}

\acknowledgments
 The authors thank S.~Aoki, H.~Fukaya, M.~Laine, M.~I.~Buchoff, and 
 N.~H.~Christ for useful discussions and correspondences.
 They are also grateful to the hospitality of 
 the Central China Normal University
 during the workshop ``High Energy Strong Interactions:~
 A School for Young Asian Scientists,'' 
 where part of this work has been carried out.
 This work was supported by the RIKEN iTHES Project, and JSPS KAKENHI 
 Grants No. 25887014 and No. 26887032.


\begin{thebibliography}{99}
\bibitem{reg}
  An ultraviolet regularization preserving chiral symmetry is assumed. 
  An example is the lattice regularization with overlap fermions. 
  However such details are not important for subsequent analysis.
\bibitem{B}
  Note that, in two-flavor QCD in a magnetic field, the symmetry ${\cal G}$ is 
  explicitly broken to $\U(1)_{\rm V3}\times \U(1)_{\rm A3} \times \UA$, for which
  one can repeat a similar argument. Here $\U(1)_{\rm V3}\times \U(1)_{\rm A3}$ symmetry
  is a vectorial and axial rotation generated by the third component of isospin.
\end{thebibliography}
\end{document}